# Spatially resolved determination of the electronic density and temperature by a visible spectro-tomography diagnostic in a linear magnetized plasma

V. Gonzalez-Fernandez*[1], P. David[2], R. Baude[3], A. Escarguel[1], Y. Camenen[1]

[1] *Aix Marseille University, CNRS, PIIM UMR 7345, Marseille, France*
[2] *Max Planck Institute for Plasma Physics, Boltzmannstr. 2, 85748 Garching, Germany*
[3] *APREX Solutions, campus ARTEM, 2 allée A. Guinier 54000 Nancy, France*

E-mail: veronica.gonzalez-fernandez@univ-amu.fr, alexandre.escarguel@univ-amu.fr

**Abstract**

In this work, a non-intrusive, spatially resolved, spectro-tomographic optical diagnostic of the electronic density and temperature on the linear plasma column Mistral is presented. Coupling of spectroscopy and tomography technique gives access to the local plasma light emission at different wavelengths (visible and near IR) in an argon plasma. Taking advantage of the symmetry of the Mistral experiment, the diagnostic results are validated and the plasma is found to correspond to a corona equilibrium state. With the use of another spectrometer and a Langmuir probe, we propose a non-intrusive method to determine the electronic density and temperature of each pixel of the tomographic images of the plasma. The obtained results are in good agreement with the Langmuir probe ones.

1. **Introduction**

Multiple applications in different areas such as surface treatment or etching processes, plasma engines or controlled fusion, makes plasmas' diagnosis a topic of high relevance [1-4]. The measurements of fundamental parameters such as the electronic temperature and density are essential to understand the physics of complex plasmas. Both parameters have been classically measured with different kinds of probes [5-7], but they can significantly disturb the plasma.

For that reason, other methods have been proposed, as for example emission spectroscopy [8, 9]. This technique allows the non-intrusive determination of a wide range of plasma parameters such as electron, ion, neutral and metastable temperatures and densities along a single line of sight (LoS). The problem of these classical methods is the lack of spatial resolution, due to the fact that the signal measured corresponds to the sum, over the LoS length, of the local emissivity distribution of the plasma.

Tomography diagnostics are widely used, e.g. computed tomography scan in medicine, because it is a non-intrusive technique that can provide good spatial resolution. The tomographic technique has been applied to plasma diagnostics since the last two decades, often with the aim of accurately reconstructing the structure of the plasma [10, 11].

Spectro-tomography combines the advantage of these two approaches to simultaneously offer spatial and spectral resolution. This method provides 2D maps of light intensities at different wavelengths [12], giving access to the plasma electronic density and temperature.

The experimental set-up and the tomographic method are presented in section 2 and 3, respectively. The experimental results are detailed in section 4 with 3 sub-sections:

- Experimental confirmation of the corona equilibrium in Mistral,
- Validation of the spectro-tomographic results,

- 2D measurement of $n_e$ and $T_e$ in Mistral.

## 2. Experimental setup

The linear magnetized plasma device Mistral [13] (Figure 1) is used for the fundamental study of transport in cold plasmas ($T_e \sim 3$ eV and $T_i \sim 300$ K). The plasma is a thermionic discharge in argon, created by 30 to 40 eV primary electrons emitted by 32 tungsten filaments, with a typical pressure of P ≈ $10^{-2}$ Pa in the source chamber and an electronic density $n_e \approx 10^{15}$ m$^{-3}$. The Mistral's vacuum chamber is a 1.2 m long and 0.4 m wide cylinder with a typical magnetic field of 0.016 T.

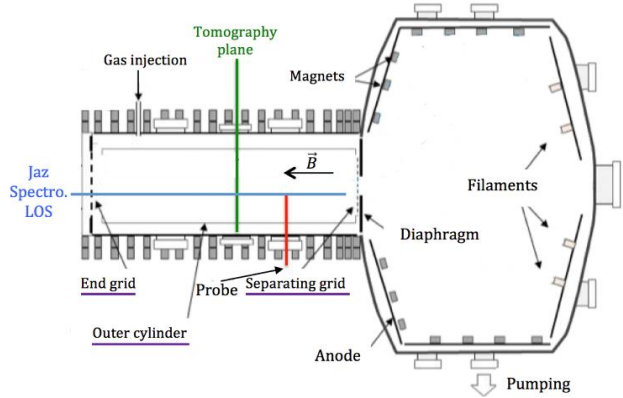

*Figure 1: The Mistral linear plasma device. The purple underlined elements can be individually polarized, controlling the longitudinal (end and separating grid) and the radial (outer cylinder) boundary conditions of the plasma. The size is 1 m length, Φ = 1.4 m for the source chamber and 1.2 m length and Φ = 0.4 m for the linear plasma chamber.*

The plasma is limited by a grid on each end of the column and an outer cylinder at r = 10 cm. The two grids and the cylinder can be independently polarized between -60 and 60 V. The magnetic field and different polarized surfaces (underlined on Figure 1) in both the source and linear plasma chamber allow for a stable and reproducible control of the plasma state during several hours.

The spectro-tomography diagnostic (see Figure 2) consists of two main parts: a tomographic device for the acquisition of 49 crossed LoS and an emission spectrometer. Two large windows, on the side and on the top of the linear chamber allow for an easy optical access of the plasma (shown as the *tomography plane* in Figure 1). 49 optical fibres (core diameter 200 μm) are positioned outside of the vacuum chamber, in front of these windows [14]. Each fibre facing the plasma is placed at the focal plane of a micro-lens (3.4 mm in diameter and 9.85 mm focal length), collimating the lines of sight. The micro-lenses are mounted on individual mechanical holders. The fibres are divided in two fans, as shown in Figure 2. The fan distribution is known to lessen the effect of noise by providing a wide range of viewing angles [15]. This optomechanical system has been carefully aligned. Indeed, because of the size of the collimation system and the distance to the plasma, a small misalignment can lead to a sizeable displacement of the line of sight. To well-adjust the collimation and position of the fibres, a laser has been retro-injected in the bundle and projected at 1 m length, a distance similar to the lens plasma distance.

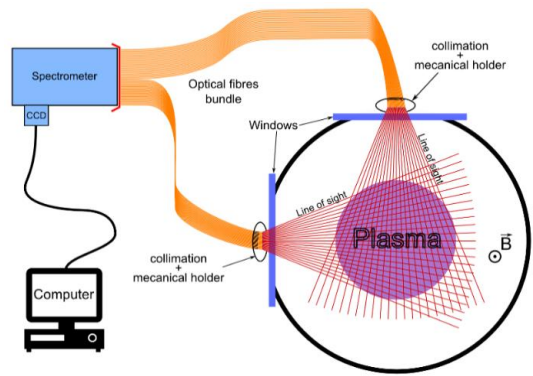

*Figure 2: Schematic of the spectro-tomography diagnostic installed on Mistral.*

On the other side of the bundle, the 49 fibres face the entrance slit of the Princeton Instruments Isoplane 160 imaging spectrometer that provides high resolution across the entire focal plane. The spectrometer has three diffraction gratings (150, 600 and 1200 lines/mm) and an adjustable entrance slit with an 11 mm effective height. The measurements shown in this work have been done with the 600 lines/mm diffraction grating and the entrance slit width fixed at 50 µm. A PROEM-HS, Princeton Instruments EMCCD camera (1024 x 1024 pixels, pixel size 13.6 x 13.6 µm) is positioned at the image focal plane of the Isoplane spectrometer. For this work, 49 Regions of Interest (ROI) (each of 1024 pixels length, and 13 pixels height) evenly spaced image the 49 LoS. Measurements were taken at different central wavelengths (from 400 nm to 900 nm, 50 nm steps), to cover the whole visible spectrum and the near infrared. For a better signal to noise ratio, each spectrum is averaged over 20 successive acquisitions, with 650 ms exposure time as maximum. A complete acquisition lasts approximately 2 minutes whereas the plasma in Mistral is stable during several tens of minutes. The spectral response of the diagnostic is obtained with a calibrated black-body applied to the Isoplane spectra. The black-body is an integrating sphere (Lambertian source) that allows transforming the arbitrary units (counts) from the spectrometer to plasma emissivity (µW cm$^{-2}$ sr$^{-1}$ nm$^{-1}$). This calibration remove any ulterior dependence from the instruments.

The spectro-tomography results are validated by comparing them with the spectra measured with a low spectral resolution spectrometer (JAZ-Ocean Optics). This spectrometer has three diffraction gratings centred on three complementary parts of the visible spectrum and near infrared, providing a whole measurement range from 400 to 980 nm. The JAZ line of sight is composed of a 600 µm trifurcated optical fibre placed at the focal point of a f'=200 mm lens. The resulting diameter of the line of sight, one meter from the lens, is 8 mm. The JAZ line of sight is placed at the end of the plasma column, parallel to the magnetic field lines of the Mistral solenoid (see "JAZ spectro LoS" on Figure 1). The JAZ spectrometer was calibrated following the same black-body protocol as the Isoplane spectrometer. In addition, a Langmuir probe (Scientific Systems, Smartprobe) radially moving from the plasma centre to its edge allows an independent measurement of $n_e$ and $T_e$ [16]. As can be seen in Figure 3, the logarithmic plot of a typical probe characteristic in Mistral is approximately linear between 10 V and 18 V. Then, the electron temperature can be deduced from the inverse of the slope of the characteristic in that bias region. The whole probe characteristic procedure for $n_e$ and $T_e$ measurement is detailed in reference [16].

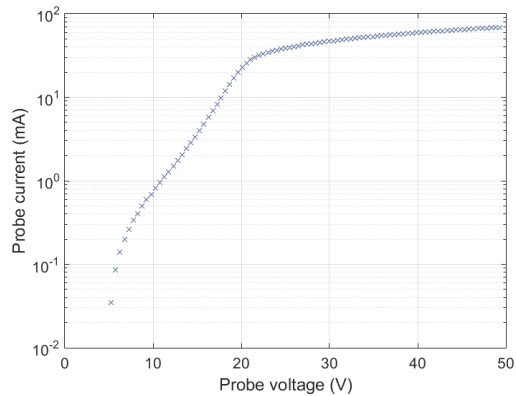

*Figure 3: Typical Langmuir probe characteristics for Mistral, presented with a semi logarithmical scale. The electron temperature is calculated from the inverse of the slope in the linear part.*

## 3. Tomographic inversion

The tomographic diagnostic has been explained in detail and validated in previous works [14, 17], so only the most important features will be referenced here. The numerical tomographic inversion is based on a finite-element scheme. With i=1,…,L intersecting LoS, the tomographic technique gives access to the local plasma emissivity divided into pixels, through the tomographic relation [18]:

$$f_i = \sum_i \sum_j T_{ij} g_j \qquad (1)$$

The measured signal $f_i$ on the i$^{th}$ line of sight is linked to the local emissivity $g_j$ of the plasma in the j$^{th}$ pixel, through the transfer matrix $T_{ij}$. Inverting the transfer matrix gives access to the plasma emissivity. The discretisation problem leads to an ill-conditioned transfer matrix. Consequently, a careful procedure is needed for the inversion process. Instead of directly solving the system, we search for the local emissivity $g_j$ through the minimisation of $\phi = \frac{1}{2}(T_{ij}g_j - f_i)^2 + \alpha R$ with $\alpha$ a positive weighting parameter and $R$ a regularizing functional. Several methods are available to obtain the best regularisation [19]. The optimization of the tomographic inversion has been thoroughly discussed in a previous work [20]. From this, we chose the alpha parameter equal to 0.1 and a second order regularization. No hypothesis has been made on the plasma shape or position.

The following procedure has been applied to the 49 raw spectra: first, the instrumental function obtained from the black-body calibration is applied to the spectra. Then, the wavelength integrated intensities of the most intense emission lines in each ROI are calculated. Finally, the tomographic reconstruction is applied to each emission line.

## 4. Experimental results

In this section, we first focus on the equilibrium model corresponding to the plasmas in Mistral. Secondly, the spectro-tomography results are experimentally validated. Finally, an example of 2D measurements of $n_e$ and $T_e$ in Mistral is presented.

- *Corona model*

The corona model is a simple description of the distribution of the excited atomic/ionic levels populations. The population of an excited state results from the balance of electron impact excitation from the ground state and decays by spontaneous emission [21]. The plasma can be described by the corona model if the distribution of the excited states populations can be fitted with the following expression with x ≤ 3 [22]:

$$\frac{N_i(p_i)}{g_i(p_i)} \alpha \, p_i^{-x} \qquad (2)$$

where $N_i$ and $g_i$ are the population density of the excited levels and the degeneracy of the excited level, respectively. The effective principal quantum number $p_i$ of the excited states is given by [22]:

$$p_i = \sqrt{\frac{E_H}{E_\infty - E_i}} \qquad (3)$$

with $E_H$ the Rydberg constant (13.6 eV), $E_\infty$ the ionization energy of the considered species and $E_i$ the energy of the excited level *i*. The normalized distribution of the excited population of Ar I has been measured in Mistral with the JAZ spectrometer. As can be seen in Figure 4, $N_i/g_i$ is in good agreement with a $p_i^{-3}$ fit. Following the classical expression of the optical depth of a Doppler broadened emission line [21], the Mistral plasma is optically thin.

The 2D map of $n_e$ and $T_e$ can be deduced from the argon emission lines intensities obtained by spectro-tomography if the relation between them is known. Taking advantage of the ($n_e$, $T_e$) homogeneity along the JAZ spectrometer line of sight, we have acquired a series of JAZ spectra and measured $n_e$ and $T_e$ with a Langmuir probe radially positioned into the JAZ LoS.

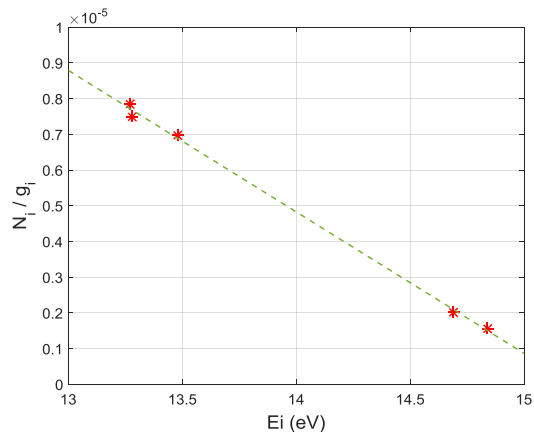

*Figure 4: Fit of the expression (3) with x=3 for the excited levels of neutral argon in Mistral.*

The measurements were performed in an argon plasma, with the following configuration: $P_{Ar}$=0.054 Pa, B=16 mT, grounded collector, floating separating grid and non-polarised outer cylinder.

The electronic density has been progressively changed without a strong variation of the electronic temperature. This is possible by increasing/decreasing the voltage current applied to the 32 filaments of Mistral, giving rise to an increase/decrease of the ionising primary electron flux incoming in the plasma column from the source, and then to a corresponding electronic density variation. Figure 5 shows the intensity of several Ar I wavelengths (693, 696, 750, 751 and 842 nm) when the voltage of the filaments are modified from 16.6 to 11.4 V, leading to an electronic densities decrease from $1.34 \cdot 10^{11}$ to $6.58 \cdot 10^{8}$ cm$^{-3}$ with a fixed pressure of 0.1 Pa. At the same time, $T_e$ decreases from 2.7 to 1.7 eV. These lines where selected for different reasons: the ones at 750 and 751 nm correspond to excited levels populated directly from the ground level. The Ar I emission lines at 693, 696 and 842 nm are sufficiently intense to be detected even when the electronic density is low. All the lines plotted in Figure 5 show a linear trend. This is coherent with the corona model, namely the intensity of a neutral line is directly proportional to the electronic density.

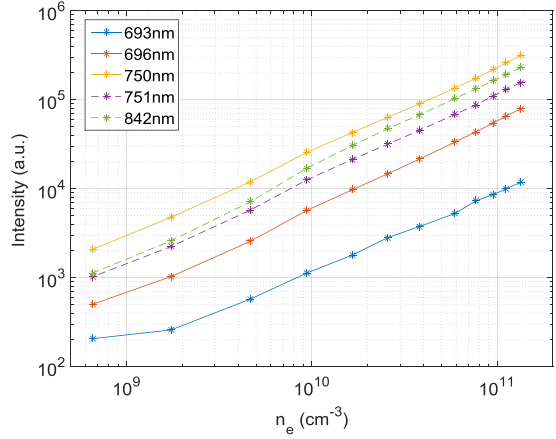

*Figure 5: Intensity of Ar I lines vs. the electronic density $n_e$ in an argon plasma in Mistral, showing a linear dependence.*

Although the plasma emission is dominated by neutral argon lines [16], the spectrum recorded by the JAZ spectrometer has also several lines corresponding to ionized argon (Ar II). In the frame of the coronal model, their intensities are directly proportional to $n_e^2$.

Figure 6 shows, the $n_e$ dependence of the intensities of two Ar II emission lines. As can be seen, linear fits of the experimental data are in good agreement for $n_e > 10^{10}$ cm$^{-3}$.

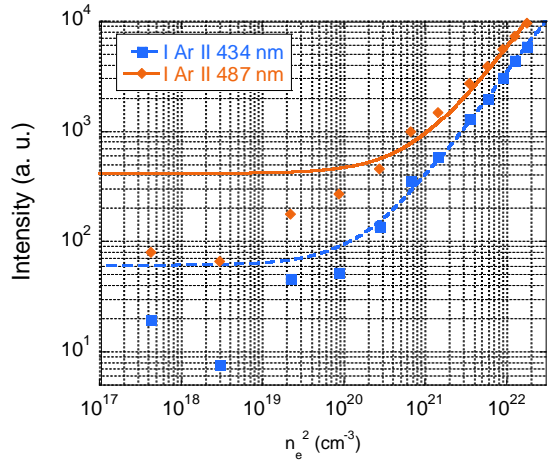

*Figure 6: Intensity of two Ar II lines vs. the square of the electronic density in an argon plasma in Mistral, showing a linear dependence for $n_e > 10^{10}$ cm$^{-3}$.*

In conclusion, we have experimentally checked the following relations for the intensities $I_{ArI}$ and $I_{ArII}$ of neutral and ionized emission lines, respectively:

$$I_{ArI} = n_e \cdot f_1(T_e) \quad (4)$$

$$I_{ArII} = n_e^2 \cdot f_2(T_e) \quad (5)$$

It is important to note that the relation (5) is experimentally found to be valid only for $n_e > 10^{10}$ cm$^{-3}$. However, as we will see in part 4.3.2, the electron density range in the Mistral plasma column is always larger than this value. This is an important step to measure ($n_e$, $T_e$) by spectro-tomography. The electronic temperature dependence of the emission lines intensities, $f_1(T_e)$ and $f_2(T_e)$ (shown in Figure 7) depend mainly on the excitation rate by electronic collisions from the fundamental level. Two populations of electrons can populate the excited levels: the secondary electrons and the few percent of primary electrons ($n_{pe}/n_{se} \approx 0.03$) [16]. The primary electrons are in the minority, but can play an important role because of their relatively high energy.

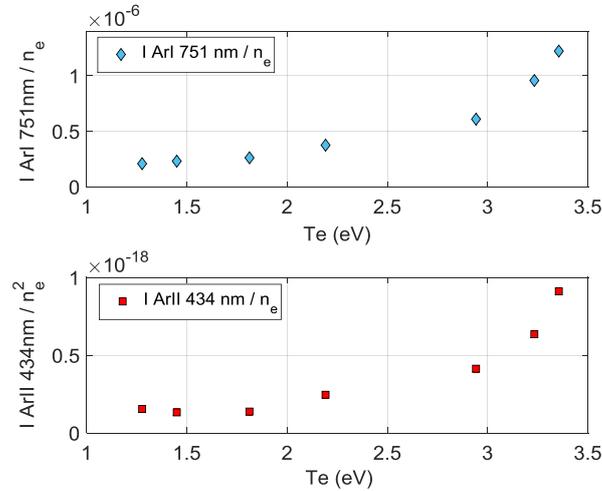

Figure 7: Up: $f_1(T_e)$, resulting from the ratio of IArI (751nm) / $n_e$. Down: $f_2(Te)$, resulting from the ratio of I ArII (434nm) / $n_e^2$.

- **Validation of the spectro-tomography technique**

As the plasma parameters are homogeneous along the magnetic field lines, it is possible to validate the results of the spectro-tomography diagnostic: the normalized radial distributions of the argon emission lines intensities extracted from the 2D images must correspond to the results obtained with the JAZ spectrometer. Figure 8 shows the Ar I (751 nm) emission line radial distributions in an argon plasma with the same experimental configuration as the previous section. As can be seen, the two sets of data show similar radial profiles for the normalised intensities. Experimental fluctuations of the intensity of the JAZ spectrometer have been found to be equal to 10%. In the case of spectro-tomography, the error bar has been considered as the one due to the tomography inversion [14]. The horizontal error bars represent the size of a pixel. In our case, with 7x7 pixels matrix reconstruction, a pixel size corresponds to 15.7 mm.

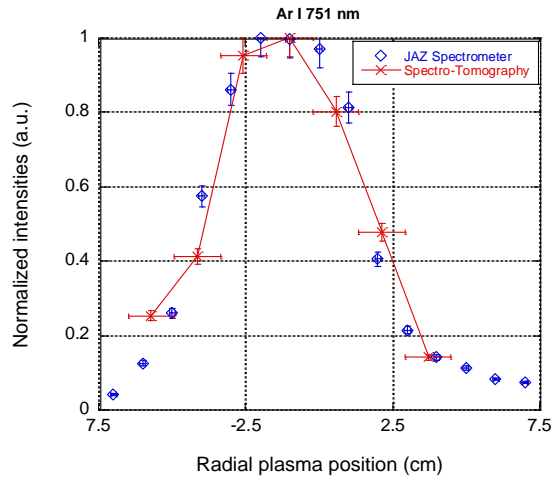

Figure 8: Normalized radial profiles of the Ar I (751 nm) emission line in an argon plasma, obtained with the JAZ spectrometer (blue diamonds) and extracted from 2D images of the spectro-tomography diagnostic (red line).

- **Calculation of $T_e$ and $n_e$ by spectro-tomography**

A classical approach for the realisation of an electron density/temperature diagnostic by emission spectroscopy is to compare some chosen emission lines intensities ratio to an atomic physics model. For example, the measurement of $T_e$ and $n_e$ in edge plasmas of tokamaks is usually done by the use of three He I emission lines ratios coupled to a collisional-radiative code [23].

This method has been applied in Mistral for helium plasmas and the high precision collisional-radiative code SOPHIA developed by F. B. Rosmej, taking into account particle diffusion and suprathermal electrons beam [24]. This work showed how complex is the atomic physics in Mistral plasmas. Then, considering that the atomic physics data of neutral/ionized argon is much less complete than helium ones, we chose not to develop an argon collisional-radiative model. Instead, a method based on spectroscopic calibration with Langmuir probe acquisition for ($n_e$; $T_e$) measurement coupled to spectroscopic acquisition of argon spectra was developed. This procedure must be operated only once.

In this section, we present how to measure a 2D map of $n_e$ and $T_e$ with a purely optical diagnostic for a corona equilibrium plasma, after a probe calibration phase. For the experimental conditions detailed in Section 2 with 49 lines of sight, we perform the tomographic inversion on 49 = 7 x 7 pixels (size 16 x 16 mm). As mentioned before, the real potential of this technique is the spectral resolution: the capability to perform simultaneously the tomographic reconstructions of emission lines at different wavelengths in the whole range of the visible spectrum. In Figure 9, three tomographic reconstructions are shown, for three different wavelengths: 416, 549 and 751 nm. The pixels at the edge of the domain can present a higher noise level than the central ones, and some of them present a zero value.

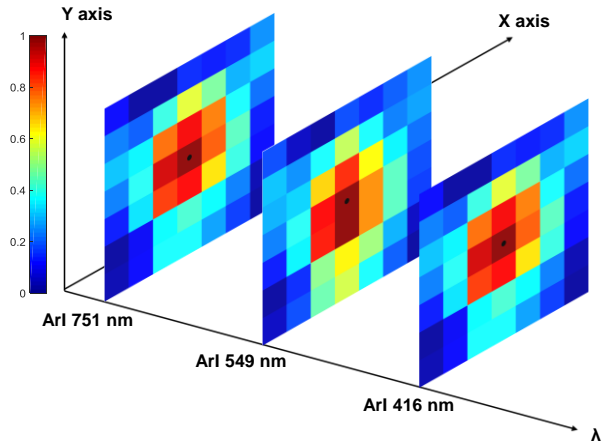

*Figure 9: Tomographic reconstructions for three different Ar I wavelengths at: 416, 549 and 751 nm, with normalized intensities.*

- *4.3.1 Determination of Te*

We have checked that the plasma in Mistral corresponds to the corona equilibrium. Then, using equations (4) and (5), we have:

$$\frac{I_{Ar\,II}}{(I_{Ar\,I})^2} = \frac{f_2(T_e)}{f_1(T_e)^2} \qquad (6)$$

Which is independent of $n_e$. The ratio of two neutral Ar lines would be as well independent of $n_e$. However, the upper states are populated in a similar way so that such ratio is poorly dependent of $T_e$.

JAZ spectrometer and probe acquisitions were simultaneously performed along the radial profile of the plasma (with fixed plasma parameters), to measure the $T_e$ dependence of the ratio $I_{ArII}$ (434 nm) / ($I_{ArI}$ (751 nm))$^2$. The result is shown in Figure 10, with a linear fit of the data. These specific wavelengths have been chosen because of the good signal-to-noise ratio.

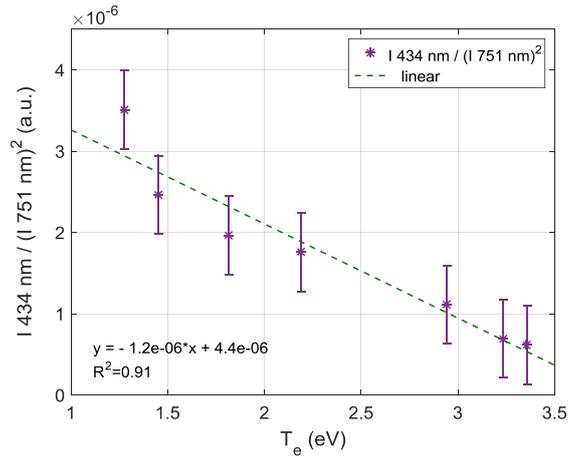

*Figure 10: Intensity of an ionized neutral line divided by the squared intensity of a neutral argon line: $I_{ArII}$(434 nm) / $I_{ArI}$(751 nm)$^2$ vs. the electronic temperature. A linear fit is applied.*

Figure 11 shows an example of measurement of this ratio by spectro-tomography for the 7 x 7 matrix of pixels. Some of the edge pixels of the ratio $I_{ArII}$ (434 nm) / $(I_{ArI}$ (751 nm))$^2$ present Not-a-Number values because in the original tomographic reconstructions it is possible to find pixels with zero value. To avoid non-realistic values of $n_e$ and $T_e$, a threshold has been imposed: all the pixels that not reach the 25 % of the maximum value of $I_{ArII}$ (434 nm) and the 15 % of $I_{ArI}$ (751 nm) have been discarded; and represented by grey pixels in Figure 11.

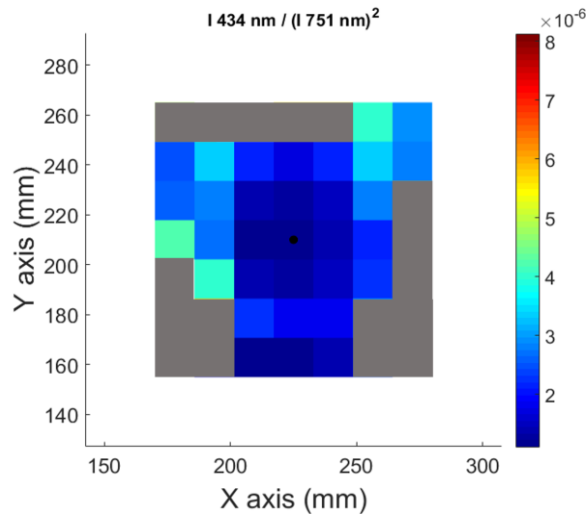

*Figure 11: Intensities' ratio of $I_{ArII}$(434 nm) / $I_{ArI}$(751 nm)$^2$ obtained by the division of the corresponding tomographic reconstruction, pixel by pixel.*

The 2D electronic temperature map is presented in Figure 12. The value of $T_e$ in each pixel has been calculated combining the values of the ratio $I_{ArII}$(434 nm) / $(I_{ArI}$ (751 nm))$^2$ of each pixel in Figure 11 and solving the equation presented in Figure 10. The electronic temperature in the plasma centre is nearly constant, with a value around 3 eV. $T_e$ decreases through the plasma limits, reaching low values around 1.5 eV.

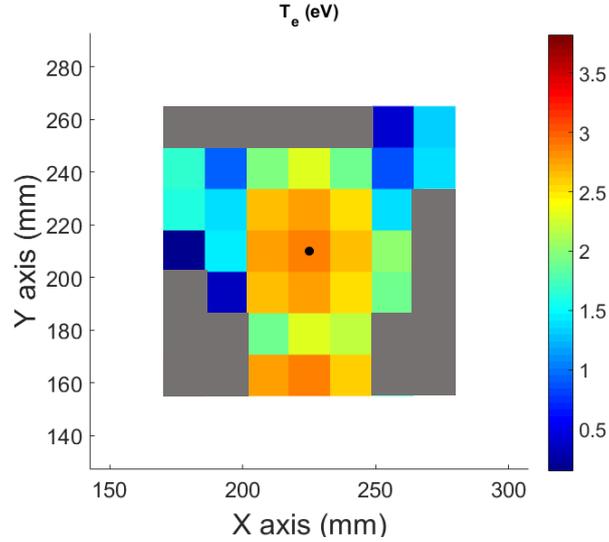

*Figure 12: 2D electronic temperature map measured by spectro-tomography.*

Figure 13 shows the comparison of $T_e$ values deduced from the Langmuir probe characteristics and from the fit of Figure 10 applied to a row of pixels of the tomographic inversion, matching the location of the horizontally movable Langmuir probe. The graph shows only three points corresponding to the path of the probe (central row of the tomographic reconstruction). The fourth pixel has been omitted because it provides a negative value of $T_e$. Indeed, the plasma is too disturbed if the probe and its holder go further away than the plasma centre. The two data series are in very good agreement, with a maximum difference of about 7 %. The standard deviation error bars for $T_e$ measured with the Langmuir probe is 20%. The error bars of $T_e$ obtained with the spectro-tomography diagnostic have been deduced from the tomographic inversion ones. The horizontal errors of $T_e$ obtained by probe and spectro-tomography refer to the probe tip length and the pixel length, respectively.

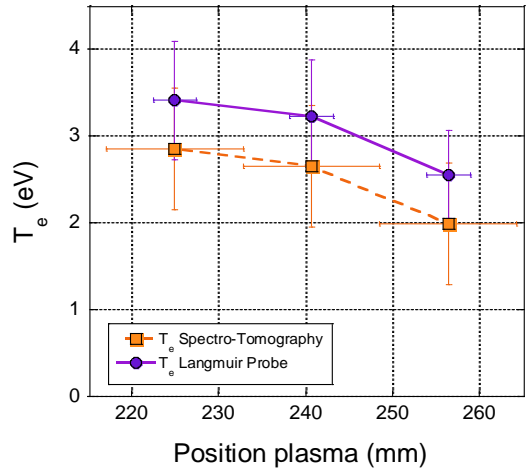

*Figure 13: Comparison of the electronic temperatures obtained by the tomographic measurements (orange sqaures) and with the Langmuir probe (purple dots).*

- **4.3.2 Determination of $n_e$**

In this section we describe the method to measure the electronic density for each pixel of the tomographic reconstruction, with the use of the measured values of $T_e$. In order to provide a method as robust as possible, we are considering again the ratio of two lines: $I_{ArII}$ (434 nm) / $I_{ArI}$ (751 nm). A lower precision is expected, as it is a more indirect method than for the measurement of $T_e$. By considering the ratio $I_{ArII}$ (434 nm) / $I_{ArI}$ (751 nm), the $f_3(T_e)$ function can be expressed as follows:

$$f_3(T_e) = \frac{f_2(T_e)}{f_1(T_e)} = \frac{I_{Ar\,II}/I_{Ar\,I}}{n_e} \quad (7)$$

In the same way as in the precedent section, the electron temperature dependence of $f_3(T_e)$ has been measured by coupling the JAZ spectrometer and the Langmuir probe diagnostic. Figure 14 shows the linear fit applied to $f_3(T_e)$.

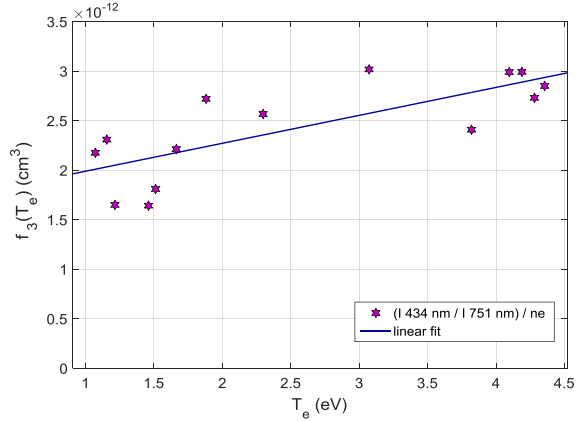

*Figure 14: $f_3(T_e)$ calculated with the JAZ spectrometer and the Langmuir probe diagnostic. A linear fit is applied.*

Following the same procedure, Figure 15 shows the intensities ratio matrix of the emission lines Ar II (434 nm) and Ar I (751 nm). As in the previous case, some pixels give non-realistic values.

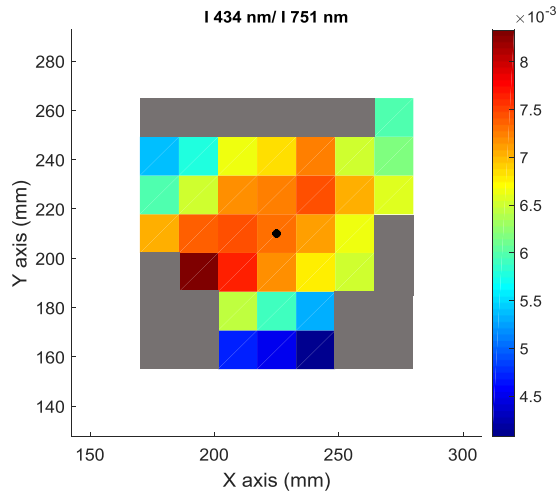

*Figure 15: Intensities ratio of the emission lines Ar II (434 nm) / Ar I (751 nm), measured by spectro-tomography.*

Electronic density for each pixel is computed from the equation (8) by using the values of $T_e$ already found by spectro-tomography and the fit of $f_3(T_e)$. The results are shown in Figure 16. At the plasma centre, the mean electronic density is $6.9 \cdot 10^{10}$ cm$^{-3}$, in the same range than the mean electronic density measured with the Langmuir probe equal to $5.6 \cdot 10^{10}$ cm$^{-3}$. A slight decrease of $n_e$ can be observed for increasing radius, except for some noisy pixels at the periphery.

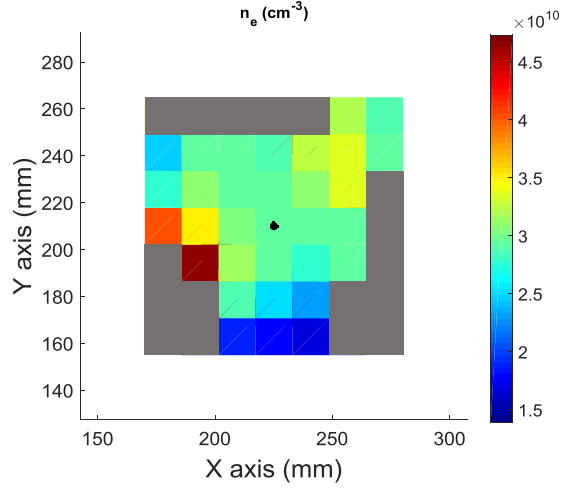

*Figure 16: 2D electronic density map measured by spectro-tomography.*

To summarize, the following procedure is applied for each pixels of the spectro-tomographic inversion:

- The value of $T_e$ is deduced from the experimental ratio $I_{ArII}(434\ nm)/I_{ArI}(751\ nm)^2$ and the linear fit presented in Figure 10.
- Then, the value of $n_e$ is obtained from the relation $n_e = (I_{ArII}(434\ nm)/I_{ArI}(751\ nm)) / f_3(T_e)$ and the value of $T_e$ obtained previously.

- *4.3.3 Determination of* $n_e$: an alternative way

It is possible to calculate $n_e$ with an alternative method described in this part. As it has been shown in Equation (4), the intensity of a neutral Ar line is directly proportional to $n_e$ and to a function of the temperature. Then, the relation can be expressed as follow:

$$n_e = \frac{I_{Ar\ I}}{f_1(T_e)} \qquad (8)$$

So, the intensity of a neutral Ar line divided by $f_1(T_e)$ measured as shown in the previous part allows a direct measurement of the electronic density. The resulting $n_e$ map is shown in Figure 17.

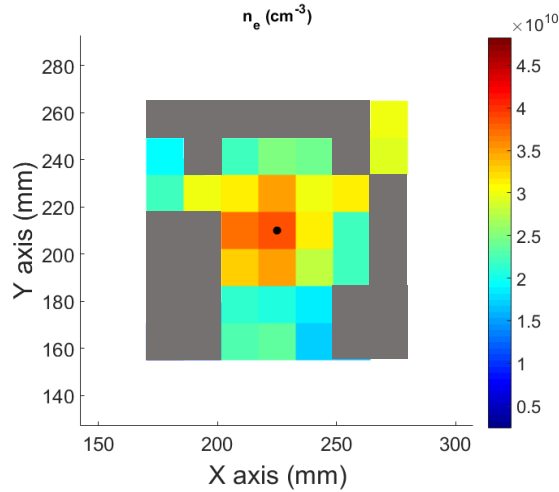

*Figure 17: 2D electronic density map measured by spectro-tomography, (alternative method).*

Therefore, for this alternative method, the following procedure is applied for each pixels of the spectro-tomographic inversion:

- The value of $T_e$ is deduced from the experimental ratio $I_{ArII}(434\ nm)/I_{ArI}(751\ nm)^2$ and the linear fit presented in Fig. 10.
- Then, $f_1(T_e)$ is deduced from Fig. 7 and the value of $n_e$ is obtained from the ratio $I_{ArI}/f_1(T_e)$.

The comparison of the measurement of ne by spectro-tomography and by the Langmuir probe installed on the Mistral experiment are shown on Figure 18. The blue diamonds represent the values of $n_e$ measured with the Langmuir probe, meanwhile the red dots/green triangles show the values of $n_e$ obtained with the first/second (alternative) methods, respectively. The two spectro-tomography methods used for measuring the electronic density are in good agreement. The more direct alternative method allows a better reproduction of the $n_e$ spatial variations than the first one. However, the tomographic measurements are found to be lower than the probe values. One should note that the probe is strongly perturbative for the plasma, then the Langmuir probe cannot be an absolute reference for the comparison. The horizontal uncertainties are the same that in the $T_e$ case. The ne standard deviation error bar obtained with the Langmuir probe is 10 %.

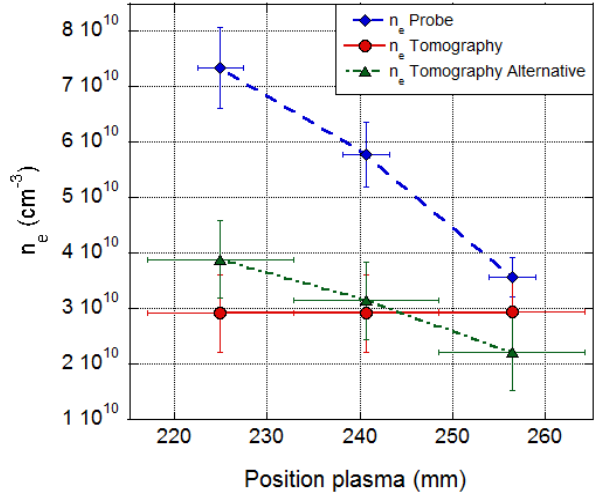

*Figure 18: Comparison of the electronic temperatures obtained by the first tomographic method (red dots), the alternative tomographic method (green triangles) and with the Langmuir probe (blue diamonds).*

## 5. Summary

In this work, a visible spectro-tomographic diagnostic installed on the Mistral experiment is presented. It is an update of a previous tomographic setup [14,17]. In the present case 49 optical fibres, divided in two fans, collect the plasma emission and send it to an imaging spectrometer, giving a simultaneous access to the brightness of emission lines in the whole visible and the near IR argon spectrum. Specifically developed software operates the tomographic reconstruction [14,17], leading to 7 x 7 pixels plasma images, with 16 mm square pixels.

With the help of a second spectrometer facing the plasma column and a Langmuir probe, taking advantage of the plasma symmetry, the tomographic reconstruction is validated by the comparison of radial profiles of intensity of emission lines. Moreover, the corona equilibrium is experimentally checked for the plasmas conditions in Mistral.

The powerful ability of the spectro-tomographic diagnostic to acquire simultaneously 2D integrated intensity of several emission lines allows to access to a wide kind of physical parameters. In the case of Mistral, we show how it is possible to measure 2D maps of the electronic density and temperature for plasma in corona equilibrium. Taking into account the ratio of ionized and neutral argon lines, we obtain spatial distribution of the electronic density and temperature, in good agreement with Langmuir probe results. The precision of the results is mainly limited by the precision of the tomographic process and by the relatively low number of LoS.

The goal of this work is to show that spectro-tomography is a reliable diagnostic allowing the measurement of important plasma parameters by purely optical methods. In the future, two strategies are planned to improve the diagnostic: first, we have begun the realisation of a transmission imaging spectrometer with an entrance slit height larger than 20 mm. This would allow increasing significantly the number of optical fibres illuminating the entrance slit, and consequently the number of LoS. Second, we are evaluating the possibility of realizing optical tomography with mirrors surrounding the plasma, allowing several views of the plasma. This would increase considerably the numbers of crossing views of the plasma.

**Acknowledgements**

This work has been carried out within the framework of the French Federation for Magnetic Fusion Research and of the Eurofusion consortium, and has received funding from the Euratom research and training programme 2014-2018 and 2019-2020 (work package WPEDU) under grant agreement No 633053. The views and opinions expressed herein do not necessarily reflect those of the European Commission.

**References**

1. Lochte-Holtegreven, W. Plasma-Diagnostics. *Ed. North-Holland, Amsterdam* (1968)
2. Hieftje, G. M. Plasma diagnostic techniques for understanding and control. *Spectrochim. Acta B*. **47**, 3-25 (1992)
3. Adamovich, I., Baalrud, S. D., Bogaerts, A. et al. The 2017 Plasma Roadmap: Low temperature plasma science and technology. *J. Phys. D: Appl. Phys*. **50**, 323001 (2017)
4. Samukawa, S., Hori, M., Rauf, S. et al. The 2012 plasma roadmap. *J. Phys. D: Appl. Phys.* **45**, 253001 (2012)
5. Cherrington, B. E. The use of electrostatic probes for plasma diagnostics-A review. *Plasma Chem. Plasma Process.* **2**, 113-140 (1982)
6. Hopkins, M. B. & Graham, W. G. Langmuir probe technique for plasma parameter measurement in a medium density discharge. *Rev. Sci. Instrum.* **57**, 2210-2217 (1986)
7. Hershkowitz, N. & Cho, M. H. Measurement of plasma potential using collecting and emitting probes. *J. Vac. Sci. Technol.: Part A*. **6**, 2054-2059 (1988)
8. Tian-Ye, N., Jin-Xiang, C., Lei, L. Jin-Ying, L. et al. A comparison among optical emission spectroscopic methods of determining electron temperature in low pressure argon plasmas. *Chinese Phys.* **16,** 2757 (2007)
9. Van der Sijde, B. & Van der Mullen, J. A. M. Temperature determination in non-LTE plasmas. *J. Quant. Spectrosc. Radiat. Transf.* **44**, 39-46 (1990)
10. Filonin, O. V. Spectral-tomographic methods and means of studying propellant flows of ion and plasma low-thrust engines. *Optics and Spectroscopy.* **118**, 181-190 (2015)
11. Neger, T. Optical tomography of plasmas by spectral interferometry. *J. Phys. D: Appl. Phys.* **28**, 47 (1995)
12. Barni, R., Caldirola, S., Fattorini, L. & Riccardi, C. Tomography of a simply magnetized toroidal plasma. *Plasma Sci. Technol*., **20**, 025102 (2017)
13. Matsukuma, M., Pierre, Th., Escarguel, A. et al. Spatiotemporal structure of low frequency waves in a magnetized plasma device. *Phys. Lett. A*. **163**, 163-167 (2003)
14. David, P., Escarguel, A., Camenen, Y. et al. A tomography diagnostic in the visible spectrum to investigate turbulence and coherent modes in the linear plasma column Mistral. *Rev. Sci. Instrum.* **88**, 113507 (2017)
15. Decoste, R. X- ray tomography on plasmas with arbitrary cross sections and limited access. *Rev. Sci. Instrum.* **56**, 806-808 (1985)
16. Escarguel, A. Optical diagnostics of a low frequency instability rotating around a magnetized plasma column. *Eur. Phys. J. D.* **56**, 209-214 (2010)
17. David, P., Escarguel, A., Camenen, Y. & Baude, R. Characterisation of coherent rotating modes in a magnetised plasma column using a mono-sensor tomography diagnostic. *Phys. Plasmas* **23**, 103511 (2016)
18. Granetz, R. S. & Smeulders, P. X-ray tomography on JET. *Nucl. Fusion.* **28,** 457 (1988)
19. Anton, M., Weisen, H., Dutch, M. J. et al. X-ray tomography on the TCV tokamak. *Plasma Phys. Control. Fusion.* **38,** 1849 (1996)
20. Baude, R., Escarguel, A., David, P., Camenen, Y., Jones, O. M. & Meyer, O. Visible spectro-tomography: from low temperature laboratory plasmas to the WEST tokamak**.** *43rd EPS Conference on Plasma Physics* (2016)
21. Griem, H. R. Principles of plasma spectroscopy. *Academic Press, New York*. (1966)
22. Gordillo-Vázquez, F. J., Camero, M. et al. Spectroscopic measurements of the electron temperature in low pressure radiofrequency $Ar/H_2/C_2H_2$ and $Ar/H_2/CH_4$ plasmas used for the synthesis of nanocarbon structures. *Plasma Sources Sci. Technol*. **15,** 42 (2005)
23. Schweer, B., Brix. M. & Lehnen, M. Measurement of edge parameters in TEXTOR-94 at the low and high field side with atomic beams. *J. Nucl. Mater.* **266**, 673-678 (1999)
24. Lefevre, T., Escarguel, A., Stamm, R., Godbert-Mouret, L. & . Rosmej, F. B. Investigation of particle diffusion and suprathermal electrons in a magnetized helium plasma column. *Phys. Plasmas.* **21**, 023502 (2014)